# Effect of *In-vivo* Heat Challenge on Physiological Parameters and Function of Peripheral Blood Mononuclear Cells in Immune Phenotyped Dairy Cattle

S. L. Cartwright,[1] J. Schmied,[1] A. Livernois,[1,2] and B. A. Mallard[1,2]

[1]Depatment of Pathobiology, Ontario Veterinary College, University of Guelph, Guelph, Ontario Canada, N1G 2W1

[2]Centre of Genetics of Improvement of Livestock, Animal Biosciences, University of Guelph, Guelph, Ontario, Canada, N1G 2W1

[1]Shannon L. Cartwright

Tel: (519) 824-4120 ext 54655

e-mail: cartwris@uoguelph.ca

Department of Pathobiology, Ontario Veterinary College, University of Guelph, Guelph, Ontario, Canada, N1G 2W1

**ABSTRACT**

The frequency of heat waves and hot days are increasing due to climate change, which leads to an increase in the occurrence of heat stress in dairy cattle. Previous studies have shown that dairy cattle identified as high immune responders have a reduced incidence of disease and improved vaccine response compared to average and low responders. Additionally, it has been observed that when cells from immune phenotyped cattle are exposed to *in-vitro* heat challenge, high immune responders exhibit increased heat tolerance compared to average and low immune responders. Therefore, the objective of this study was to evaluate physiological parameters and the function of blood mononuclear cells in immune phenotyped dairy cattle exposed to *in-vivo* heat challenge. A total of 24 immune phenotyped lactating dairy cattle (8 high, 8 average and 8 low) were housed in the tie-stall area of the barn and exposed to an *in-vivo* heat challenge for 4 hours on 2 subsequent days, where the temperature was set at 29°C. Blood samples were taken both pre- and post-challenge each day and manual respiration rates and rectal temperatures were recorded pre challenge and every 30 minutes during the challenge. Temperature and humidity measurements were taken in correspondence with all respiration rate and rectal temperature measurements to calculate the temperature humidity index pre heat challenge and at 30-minute intervals during the heat challenge. Blood mononuclear cells were isolated from blood collected pre and post challenge and the concentration of heat shock protein 70 and cell proliferation were assessed. Results showed that average and low responders had significantly greater respiration rates compared to high responders at a temperature humidity index of 77 and above. No significant difference was observed between phenotypes for rectal temperature. High responders had a higher heat shock protein 70 concentration and greater cell proliferation after both *in-vivo* heat challenges compared to average and low responders. These results paralleled those found during *in-vitro* heat challenge adding further credence to the concept that high responders may be more resilient to heat stress compared average and low responders.

**INTRODUCTION**

Since the industrial revolution greenhouse gas emissions have been steadily increasing (IPCC, 2018), they act to trap heat in the atmosphere, not allowing it to dissipate to space. Reports suggest that recent human activity has led to a global atmospheric warming of 1°C above pre-industrial revolution temperatures and this warming is expected to continue (IPCC, 2018). In areas of larger land mass and higher latitudes atmospheric warming is even greater, warming in Canada is 2°C above pre-industrial revolution temperatures (IPCC, 2014). Elevated heat in the atmosphere is associated with numerous environmental issues and has led to the variability observed in the climate today. One impact of atmospheric warming is an increase in extreme weather events, such as heat waves, which results in an escalation in the number of hot days throughout the year (IPCC, 2014). It's difficult for livestock, but especially high producing dairy cows due to their relatively high metabolic demands, to adapt to the constantly changing and warming climate. As climate change continues, cattle will be at increased risk of both frequent and prolonged heat stress events.

Dairy cattle utilize several physiological and cellular mechanisms to manage heat accumulation during high ambient temperatures. Initially, dairy cattle will employ evaporative mechanisms increasing their respiration rate and sweating in order to remove body heat (Collier et al., 2019; Garner et al., 2016). Additionally, the heart rate will elevate, increasing blood flow and vasodilation at the periphery. This redistributes blood to the surface of the body, allowing for excess heat to be transferred to the surrounding environment (Shilja et al., 2016). Ideally these physiological mechanisms dissipate enough heat to maintain normal core temperatures. However, when the ambient temperature and humidity increase to a point where dairy cattle can no longer effectively dissipate heat, the core body temperature will increase above normal levels resulting in hyperthermia (Collier et al., 2019). Additionally, dairy cattle utilize various cellular mechanisms to protect cells from heat stress. When dairy cattle are experiencing heat stress their cells are triggered to respond similarly to other stress inducing conditions, the slightest elevation of body temperature can trigger such a response (Richter et al., 2010). The first impact of elevation of core body temperature on cells is the unfolding of intracellular proteins (Richter et al., 2010). This triggers cells to release bound heat shock proteins (**HSP**) from heat shock factor 1 (Collier et al., 2008; Jacob et al., 2017). Heat shock factor 1 forms a trimer within cells, binds the promoter region and activates transcription of heat stress genes leading to the production of various molecules involved in the stress response (Collier et al., 2008; Jacob et al., 2017). These molecules include the production of additional HSP, which act during cellular stress to repair mis-folded proteins preventing cell death (Bernabucci et al., 2010; Collier et al., 2019). Further, HSP protect proteins and organelles from damage during cellular stress (Bernabucci et al., 2010; Collier et al., 2019). These physiological and cellular mechanisms are essential for maintaining cellular homeostasis within the body during heat stress in dairy cattle.

Among livestock, dairy cattle are particularly prone to heat stress as selective breeding for increased milk production has resulted in higher metabolic heat production (Carabaño et al., 2017; Kadzere et al., 2002). Heat stress leads to numerous issues in dairy cows including reduced milk production (Nasr and El-Tarabany, 2017), impaired reproduction (Dash et al., 2016) and increased occurrence of disease (Das et al., 2016). In severe cases heat stress can even lead to mortality (Bishop-Williams et al., 2015; Das et al., 2016). Some of these issues may be a result of the impact of heat stress has on the immune system. Indeed, studies have shown heat stress impacts both the innate and adaptive immune responses. In particular, heat stress is associated with reduced complement activation (Min et al., 2016), reduced cytokine production (do Amaral et al., 2011), and impaired cellular function (Bagath et al., 2019), resulting in cattle that are more susceptible to pathogens. Additionally, heat stress has been linked to reduced immunoglobulin production and reduced cell proliferation, which may be a function of reduced cell-mediated immune response (**CMIR**) (Dahl et al., 2020). Impairment in these components of the immune system may explain the increased risk of disease and reduced response to vaccination observed during heat stress (Bagath et al., 2019). For these reasons addressing climate change should be of importance to dairy producers. Not only will climate change result in economic losses estimated between $897 million to $1.5 billion per year in the United States alone (St-Pierre et al., 2003), but climate change is expected to negatively impact the health and welfare of dairy cattle.

It has previously been shown that dairy cattle classified as high immune responders, based on their estimated breeding values (**EBV**) for immune responsiveness, have overall lower incidence of disease (Thompson-Crispi et al., 2012), improved response to vaccination (Wagter et al., 2000) and improved hoof health (Cartwright et al., 2017), while having little to no effect on production (Stoop et al., 2016). More recently it was shown that *in-vitro* heat challenge of PBMC from immune phenotyped cattle resulted in differential DNA methylation patterns that suggested an increase in the expression of genes related to cellular protection in cells from high responders compared to cells from low responders (Livernois et al., 2021). These same heat challenged PBMC also displayed increased production of heat shock protein 70 (**HSP70**) and greater proliferation from cells obtained from high responders compared to average and low immune responders (Cartwright et al., 2021). Therefore, it would be informative to evaluate the effect of an *in-vivo* heat challenge of immune phenotyped dairy cattle and the physiological and cellular mechanisms associated with heat stress. The objective of this study was to evaluate the effect of *in-vivo* heat challenge on respiration rate, rectal temperature and the function of PBMC with respect to HSP70 concentration and cell proliferation in lactating dairy cattle classified as high, average and low immune responders. It is hypothesized that high immune responding dairy cattle will have lower respiration rates and rectal temperatures when heat challenged. It was also hypothesized that cells from high immune responding dairy cattle will produce greater concentrations of HSP70 and display increased cell proliferation compared to PBMC from average and low immune responding dairy cattle after exposure to *in-vivo* heat challenge.

**MATERIALS AND METHODS**

Use of animals and experimental procedures were approved, prior to commencement of the study, by the Animal Care Committee at the University of Guelph (animal utilization protocol #3555).

*Animals*

A total of 24 (sample size calculated using one-way ANOVA in G*Power 3.1.9.7 (Faul et al., 2007), $\alpha = 0.05$, power = 0.95, standard deviation and means obtained from data evaluated in (Cartwright et al., 2021)) lactating Holstein dairy cattle previously ranked for immune response were evaluated in this study. Cattle were housed in the tie-stall area of the Elora Dairy Research Station. The tie-stall area encompasses 2 rows of stalls with 15 stalls in each row. Ventilation fans, cooling fans and heaters are evenly spaced on the ceiling of tie-stall area to allow even distribution of air flow and heat. Cattle were housed in 4 groups of 6, with each group containing 2 high, 2 average and 2 low cows. Cattle were placed randomly in the first 3 stalls of each row. Each cow was tethered within its own individual stall and had free access to water and feed through individualized water bowls and a feed bunk in front of their stall. All cattle were fed the same total mixed ration twice per day, the same ration they were fed in the lactating barn prior to being moved to the tie-stall. Additionally, cattle were milked twice daily at 4am and 4pm in their respective stalls. The cattle were comprised of various parities (ranging from parities 1-5), pregnancy status (ranging from not pregnant to 210 days pregnant) and production levels (ranging from a daily average of 20 to 44 litres per day, however when comparing between phenotypes average daily production did not differ). Each of these effects were accounted for in statistical models. Cattle were ranked for immune response using a previously described protocol (Hernandez et al., 2005; Wagter et al., 2000). Briefly, cattle were immunized i.m. with type 1 and type 2 test antigens (Wagter and Mallard, 2007). Blood was taken via the tail vein prior to immunization and 14 days after immunization. Sera was collected from blood samples and used in ELISA to assess antibody-mediated immune response (**AMIR**). Additionally, 14 days after immunization, delayed type hypersensitivity (**DTH**) tests were initiated. Triplicate measurements of skin-fold thickness were taken on each side of the tail-fold. The type 1 antigen and PBS control antigen were injected i.d. on either side of the tail fold. Twenty-four hours later, additional skin-fold measurements were taken on either side of the tail fold to assess DTH as an indicator of CMIR. Cattle were ranked as high, average, or low immune responders based on their EBV for AMIR and CMIR. For this study, 8 high AMIR-high CMIR, 8 average AMIR-average CMIR, and 8 low AMIR-low CMIR Holstein dairy cattle were evaluated.

*Heat Stress Challenge*

To create a heat challenge environment for the cattle in this study, experiments were conducted during the summer months (June-August). To validate findings from a previous *in-vitro* heat challenge, conducted with immune phenotyped dairy cattle (Cartwright et al., 2021), cattle in this study were similarly heat challenged . Initially, cattle were moved into the tie-stall area of the barn three days prior to initiation of the heat challenge (Al-Qaisi et al., 2019). During this acclimation period the thermostat remained set at the standard 10°C, cooling fans remained on and ventilation fans remained at 100% fan speed. Additionally, measured production levels of cattle stabilized within this period. Cattle were moved to the tie-stall to best ensure temperature control between groups.  On the first day prior to initiation of the heat challenge (Thermoneutral (**TN**)), rectal temperatures were taken using a digital thermometer (Life Brand) and respiration rate was measured by counting the number of breaths per minute (**min**). Additionally, blood samples were collected (using BD EDTA vacutainer tubes) via the tail vein, while cows remained their respective stalls and temperature and humidity for the tie-stall area was recorded (TN measurements were taken at 6am consistently throughout the study and THI ranged from 66 to 73, this variation was accounted for in statistical models). Blood samples were processed immediately after collection for the isolation of PBMC. After preliminary blood collection, the thermostat in the tie-stall area was increased to 29°C (the hottest temperature recorded for the lactating area of the barn), cooling fans were turned off and ventilation fan speed turned down to 10%. The tie-stall area was warmed for 2 hrs prior to the initiation of the heat challenge, which began consistently at 8am each week throughout the study (and lasted for a total of 4 hours (ending at 12pm) (Mehla et al., 2014)). Cooling fans remained off and ventilation fan speed was maintained at 10% for the duration of the heat challenge. During the 4-hour heat challenge temperature and humidity for the tie-stall area was recorded every 30 minutes to calculate temperature humidity index (**THI**) at each 30-minute interval. In conjunction with the temperature and humidity measurements, every 30 minutes throughout the 4-hour heat challenge, respiration rate and rectal temperature was measured as described above for each cow. At then end of the 4-hour challenge (**heat stress 1 (HS1)**), blood was taken from the tail vein, and returned to the lab for immediate processing and isolation of PBMC. The tie-stall thermostat was then returned to the normal optimal set point (10°C) and ventilation fan speed was turned back up to 100%. The following day, blood samples from the tail vein were taken, 18 hours after completion of HS1 treatment (**18HS1**) and the thermostat and ventilation were adjusted as described above 2 hrs prior to the commencement of heat stress 2 (**HS2**). Temperature, humidity, respiration rate and rectal temperatures were obtained during HS2 every 30 minutes as previously described for HS1. Blood samples were obtained via the tail vein at the end of HS2 and barn conditions returned to normal. On the day following HS2, final blood samples were taken via the tail vein, 18 hours after the completion of HS2 (**18HS2**). Temperature and humidity measurements collected throughout the heat stress challenge were used to calculate THI at each time point using the following formula: THI = (1.8 x temperature + 32) – ((0.55 – (0.0055 x humidity)) x (1.8 x (temperature – 26))) (Zimbelman et al., 2009). Throughout the entire heat challenge cattle were monitored to ensure they did not enter a state of extreme hyperthermia. Upon completion of the heat challenge protocol all cattle were returned to the lactating area of the barn and remained in the herd to be used for subsequent studies.

*Cell Isolation*

Peripheral Blood mononuclear cells were isolated from blood collected both pre and post heat stress challenge using a density gradient as previously described (Cartwright et al., 2021). Briefly, the buffy coat was isolated from whole blood after centrifugation (1200xg for 20 min) and diluted with PBS in a 1:2 ratio. The PBS buffy coat solution was layered over Histopaque (Sigma-Aldrich) in a 1:1 ratio and the solution was centrifuged at 1200xg for 10 min. The resulting buffy coat was isolated and washed with PBS. This solution was centrifuged at 150xg for 10 min to form a cell pellet. The supernatant was decarded and the cell pellet was re-suspended in 3 ml of PBS. The number of PBMC per ml of PBS was determined using an ORFLO cell counter (ORFLO Technologies) and cell viability was assessed by hemocytometer and trypan blue (Sigma-Aldrich). The cell numbers plated for each assay are described below.

*Heat Shock Protein 70 Concentration*

Part of the PBMC suspension obtained after each blood collection was diluted in PBS to concentration of $1.0 \times 10^6$ cells/ml. The diluted cell suspension was centrifuged at 2500Xg for 10 min to obtain a cell pellet. The supernatant was discarded, and cells were lysed by resuspending the cell pellet in a mixture of 10μl/ml of HALT protease inhibitor (Sigma-Aldrich) and M-PER Mammalian Protein Extraction Reagent (Thermo-Fisher Scientific) and centrifuging at 2500xg for 30 mins at 4°C. The liquid portion, which contained the protein, was aliquoted and stored at -80 until time of analysis. The HSP70 concentration was obtained using a commercial ELISA kit specific for bovine HSP70 (Abclonal, (Cartwright et al., 2021; Husseini, 2020; Livernois et al., 2021)) following manufactures instructions. Samples were plated in duplicate, and the average of the duplicate samples were taken and used for analysis. To ensure consistency in comparison of HSP70 concentration across all plates, additional sample aliquots were retained from one cow and assayed in duplicate on all HSP70 ELISA plates.

*Cell Proliferation*

Cells for the proliferation assay were diluted to a concentration of $5.0 \times 10^6$ cells/ml in RPMI media (Thermo-Fisher Scientific) containing 10% fetal bovine serum (Sigma-Aldrich) and Penicillin-Streptomycin (Sigma-Aldrich). Cells were plated in 6 replicates of 100 µl of cell suspension for each animal in a 96 well flat bottom culture plate (Sigma-Aldrich). Of the 6 replicates, 3 were stimulated with **ConA** (Sigma-Aldrich) mitogen at a concentration of 5µg/ml and the other 3 remained unstimulated. Additionally, 6 control replicates of RPMI media were added to each plate and PBS was added to any remaining wells to prevent evaporative loss. Plates were placed in an incubator at 37°C with 5% CO2 for 72 hours. Cell proliferation was then assessed using an MTT assay (Sigma-Aldrich) following manufactures instructions with minor modifications. Modifications included a centrifuge step, 600xg for 5 minutes, after the addition of MTT solution to sample and control wells, and another 3 hours of incubation of assayed cells at 37°C. After centrifugation the culture supernatant was discarded and MTT solvent was added to the wells. Plates were placed on a gyratory shaker for 10 min and then read using a Biotek plate reader at a wavelength of 570nm. All replicates for unstimulated and stimulated cells were averaged for each animal, with the coefficient of variation for each sample being 10% or lower. Cell proliferation for each animal was determined using the following formula: (average of stimulated cells – average of culture media) / (average of unstimulated cells – average of culture media) and presented at the stimulation index.

*Statistical Analysis*

All data were checked for normality using the Shapiro-Wilks test in R 3.6 (R Core Team, 2019). Any data not normally distributed was log transformed and re-checked for normality. Variation in respiration rate and rectal temperature between immune phenotyped cattle were evaluated using repeated measures models with an autoregressive co-variance structure in R 3.6 (R Core Team, 2019) with the following model:

$$y_{ijklmn} = \mu + i_i + t_j + l_k + p_l + m_m + g_n + e_{ijklmn}$$

where $y_{ijklmn}$ = either rectal temperature or respiration rate, $\mu$ = overall mean, $i_i$ = effect of immune response phenotype (high, average or low), $t_j$ = effect of THI at each time point (pre-heat challenge plus every 30 minutes during the heat challenge), $l_k$ = effect of parity (1, 2, 3 and ≥4), $p_l$ = effect of pregnancy status (0 = not pregnant, 1 = 1-100 days pregnant, 2 = 101-200 days pregnant, 3 = >200 days pregnant), $m_m$ = average daily production throughout study period, $g_n$ = effect of group (1,2,3 or 4) and $e_{ijklmn}$ = residual error. Variation in HSP70 concentration and cell proliferation between immune phenotyped cattle within each treatment were evaluated using general linear models in R 3.6 (R Core Team, 2019) with the following model:

$$y_{ijklmn} = \mu + i_i + t_j + l_k + p_l + m_m + g_n + e_{ijklmn}$$

where $y_{ijklmn}$ = either HSP70 concentration or cell proliferation, $\mu$ = overall mean, $i_i$ = effect of immune response phenotype (high, average or low), $t_j$ = effect of THI at time of blood sample, $l_k$ = effect of parity (1, 2, 3 and ≥4), $p_l$ = effect of pregnancy status (0 = not pregnant, 1 = 1-100 days pregnant, 2 = 101-200 days pregnant, 3 = >200 days pregnant), $m_m$ = average daily production throughout study period, $g_n$ = effect of group (1,2,3 or 4) and $e_{ijklmn}$ = residual error. Variation in HSP70 concentration an cell proliferation within phenotype across treatments were also evaluated using general linear models in R 3.6 (R Core Team, 2019) with the following model:

$$y_{ijklmn} = \mu + h_i + t_j + l_k + p_l + m_m + g_n + e_{ijklmn}$$

where $y_{ijklmn}$ = either HSP70 concentration or cell proliferation for high, average or low immune responders, $\mu$ = overall mean, $h_i$ = effect of heat challenge treatment (TN, HS1, 18HS1, HS2, 18HS2), $t_j$ = effect of THI at time of blood sample, $l_k$ = effect of parity (1, 2, 3 and $\geq$4), $p_l$ = effect of pregnancy status (0 = not pregnant, 1 = 1-100 days pregnant, 2 = 101-200 days pregnant, 3 = >200 days pregnant), $m_m$ = average daily production throughout study period, $g_n$ = effect of group (1,2,3 or 4) and $e_{ijklmn}$ = residual error. Results are presented as least squared means (**LSM**) of untransformed data. P-values are however based on normalized data with significance reported at $P < 0.05$ and trends reported at $P < 0.10$.

**RESULTS**

*Respiration Rate*

High immune responders had lower respiration rates compared to average and low responders (Figure 1). Differences in respiration rate were observed when THI reached 77, high immune responders (LSM = 46.4 breaths/min) had significantly lower respiration rates compared to average (LSM = 61.1 breaths/min, high vs average p = 0.025) and low immune responders (LSM = 61.6 breaths/min, high vs low p = 0.020). Similarly, it was observed that high responders had a significantly lower respiration rate compared to average and low responders at a THI of 78 (high LSM = 51.1 breaths/min, average LSM = 64.4 breaths/min, low LSM = 62.2 breaths/min, high vs average p = 0.042, high vs low p = 0.046) and 81 (high LSM = 51.4 breaths/min, average LSM = 66.1 breaths/min, low LSM = 73.4 breaths/min, high vs average p = 0.025, high vs low p = 0.008). A trend was observed at THI of 80, for high responders (LSM = 51.4 breaths/min) to have lower respiration rates compared to average (LSM = 63.6 breaths/min, high vs average p = 0.071) and low responders (LSM = 63.2 breaths/min, high vs low p = 0.086). It was also noted that high responders maintained a normal respiration rate (26-50 breaths per min) or close to normal range at higher THI compared to average and low responders.

*Rectal Temperature*

No significant difference was observed between the immune response phenotypes for rectal temperature for most THI values. The only significant difference observed was at a THI of 72, where high responders (LSM = 38.6°C) has significantly greater rectal temperature compared to average (LSM = 38.2°C, high vs average p = 0.010) and low responders (LSM = 38.2, high vs low p = 0.040). All phenotypes remained within normal range for rectal temperature (38.5-39.3°C) across all THI values. However, at THI of 76, there was a significant increase (p = 0.001) in rectal temperature for all phenotypes compared to lower THI values, and this rectal temperature remained elevated up to a final THI of 81.

*HSP70 Concentration*

The differences between phenotypes for the effect of *in-vivo* heat challenge on the production of HSP70 by BMC can be observed in Figure 3. Results show that PBMC from high responders (LSM = 12.1 ng/ml) produced a significantly greater concentration of HSP70 compared to both average (LSM = 9.2 ng/ml, p = 0.025) and low responders (LSM = 7.5 ng/ml, p = 0.001) after HS1. Similarly, PBMC from high responders (LSM = 12.2 ng/ml) trended towards producing a greater concentration of HSP70 compared to PBMC from average (LSM = 9.6 ng/ml, p = 0.088) and low responders (LSM = 10.3 ng/ml, p = 0.095) after HS2. It was also observed that within phenotype and between treatments, PBMC from high responders displayed a significant increase in HSP70 concentration at HS1 (p = 0.012) and HS2 (p = 0.036) compared to TN (LSM = 6.3 ng/ml), and a significant decrease in HSP70 concentration at 18HS1 (LSM = 7.4 ng/ml) and 18HS2 (LSM = 8.6 ng/ml) compared to HS1 (p = 0.006) and HS2 (p = 0.046). Conversely, PBMC from average, and low responders displayed no difference in the concentration of HSP70 produced between treatments.

*Cell Proliferation*

Results between phenotypes within treatment for cell proliferation can be observed in Figure 4. Peripheral Blood mononuclear cells from high responders (TN LSM = 1.8, HS1 LSM = 3.4, HS2 LSM = 2.8) proliferated significantly more than low responders (TN LSM = 0.8, HS1 LSM = 1.0, HS2 LSM = 0.6, for TN p = 0.043, for HS1 p = 0.028 and for HS2 p = 0.043) across all treatments. High responder PBMC trended toward greater cell proliferation compared to those of average responders (TN LSM = 1.4, HS1 LSM = 2.3, HS2 LSM = 1.1) after HS2 (p = 0.096). Similarly, PBMC from average responders trended toward greater cell proliferation compared to those of low responders after both HS1 (p = 0.084) and HS2 (p = 0.066). Within phenotype and between treatments, BMC from high responders after HS1 treatment proliferated more than TN treated cells (p = 0.023). Moreover, there was a trend toward greater cell proliferation after HS2 compared to TN (p = 0.095) for high responder PBMC. No significant difference in cell proliferation was observed between treatments for average and low responders.

**DISCUSSION**

When heat loads increase in dairy cattle, due to high ambient temperature and high humidity, internal temperature must be maintained to ensure physiological homeostasis and prevent heat stress (Polsky and von Keyserlingk, 2017). One of the ways mammals, including cattle decrease heat load is by evaporative heat loss, which can be achieved by increasing respiration rate (Polsky and von Keyserlingk, 2017). Therefore, increased respiration rate is one of the first visual clues that dairy cattle are experiencing heat stress (Collier et al., 2019; Garner et al., 2016). Typically, the normal respiration rate of adult lactating dairy cattle ranges between 26 and 50 breaths per minute (Becker et al., 2020). Therefore, a respiration rate of more than 50 breaths per minute is considered above normal and can be an early sign of heat stress (Al-Qaisi et al., 2019; Collier et al., 2019). Results from this study showed that the respiration rate of cattle remained within a normal range up to a THI of 75. Previous studies have reported that the THI range at which Holstein cattle start to experience heat stress is between 60-78. This range is dependent on the study location and production level of the cattle (Aguilar et al., 2010; Brügemann et al., 2011; Dikmen and Hansen, 2009; Hammami et al., 2013; Zimbelman et al., 2009). The respiration rates observed for cattle in this study, during heat challenge, falls within the range of what has been observed in previous studies (Lim et al., 2021; Yan et al., 2020). Where THI and respiration rates measured were indicative of mild to moderate heat stress (Mishra, 2021; Yan et al., 2020) In this study, cattle classified as low immune responders had an average respiration rate above normal when the THI reached 76. Moreover, at THI values greater than 76, both average and low immune responders had average respiration rates above normal. The respiration was significantly greater in these animals compared to high responders at a THI of 77, 78 and 81. Cattle classified as high immune responders maintained an average normal respiration rate until the THI reached 78, at which point they experienced an increased average respiration rate. However, even at THI =78, the respiration rate of high immune responders remained only slightly higher than the maximum normal value at increasing THI values (Figure 1). Taken together, these results suggest that high immune responders are more tolerant to mild to moderate heat stress compared to average and low responders.

If dairy cattle can dissipate heat effectively from their body, even during heat stress, their core body temperature will remain normal. Typically the average normal core body temperature for dairy cattle ranges between 38.5°C to 39.3°C (Fielder, 2015). Results from this study showed that all cattle maintained normal rectal temperature as THI increased and reached a maximum of 81. Additionally, rectal temperature between cattle of all immune phenotypes was similar. During heat stress it is important for dairy cattle to maintain core body temperatures to maintain internal homeostasis (Lees et al., 2019). Rising body temperature indicates ineffective heat dissipation and signals that cattle are entering a state of hyperthermia (Collier et al., 2019; Garner et al., 2016).

Although previous studies have observed hyperthermia in cattle during severe heat stress (Sammad et al., 2020), cattle in this study were able to maintain body temperature within a normal range throughout the two heat stress challenges, indicating that severe heat stress did not occur. It is likely that the THI was not high enough and/or long enough in duration to elicit overt hyperthermia. Furthermore, although still within normal body temperature range, all cattle in the study exhibited increased rectal temperature when the THI reached 76. This temperature remained constant as the THI increased to 81. This increase in rectal temperature at a THI of 76 corresponds to a similar, above normal, increase in respiration rate at a THI of 76 for low responders, and 77 for average responders. Indeed, increased respiration rate is the first sign of heat stress in cattle (Collier et al., 2019; Garner et al., 2016), and was observed here for all phenotypes, especially for average and low responders. These results concur with those observed in a study evaluating rectal temperature in dairy cattle in spring and summer months, where the authors found that although there was a significant increase in rectal temperature in summer months versus spring months, the rectal temperature for both test periods still remained within the normal range for core temperature (Rejeb et al., 2016). Rajeb et al (2016) also found significant differences in respiration rate between spring and summer months, with respiration rates in summer months being above normal.

Maintaining internal homeostasis during heat stress also requires cellular mechanisms. One of the major cellular mechanisms identified in dairy cattle in response to elevated temperatures is the production of HSP and in particular HSP70 (Bhanuprakash et al., 2016). The results from this study showed variation in the concentration of HSP70 from PBMC of high versus average and low immune responders after HS1 and HS2, where PBMC from high responders produced greater concentrations of HSP70 compared to cells from average and low responders after HS1, and trend towards greater HSP70 concentration compared to cells from average and low responders after HS2. As previously mentioned, HSP70 is produced in cells when body temperature is elevated in order to protect against and repair damaged cells to prevent apoptosis (Bernabucci et al., 2010; Collier et al., 2019; Richter et al., 2010). Therefore, these results may indicate that PBMC of high responders have a greater ability to protect themselves against damage caused by single and multiple mild to moderate heat stress events, compared to PBMC from average and low responders. Further, these results confirm those observed in a previous study evaluating the effects of an *in-vitro* heat stress challenge on PBMC from immune phenotyped cattle, where high responders produced significantly greater concentrations of HSP70 compared to average responders after one *in-vitro* heat stress challenge and a trend towards greater HSP70 concentration compared to low after two subsequent *in-vitro* heat stress challenges (Cartwright et al., 2021). Cells from high responders also displayed significantly decreased HSP70 concentrations at 18HS1 and 18HS2 compared to HS1 and HS2, respectively, which was an expected result. Indeed, it has been shown that once the threat of elevated body temperature has subsided, HSP70 will degrade its own messenger RNA to halt the production of additional HSP70 molecules (Balakrishnan and De Maio, 2006). Conversely, cells from average and low responders showed no significant difference in HSP70 concentration after the second heat challenge. One of the major issues associated with heat stress in dairy cattle is impaired cellular function (Bagath et al., 2019). Therefore, cells from average and low immune responders may be more susceptible to impairment of function as a result of heat stress, which may affect their ability to maintain defense and other functions during a heat challenge compared to high responders. Failure to effectively protect cells during heat stress could result in future health (Abdelnour et al., 2019; Kansas, 1996) and production related issues (Tao et al., 2018, 2011).

Another common problem associated with heat stress and cellular function is reduced cellular proliferation (Bagath et al., 2019). Cell proliferation is an important immunological mechanism in response to foreign pathogens. When cattle encounter an invading pathogen, cells of the immune system differentiate and proliferate to produce a robust and effective response to neutralize and eliminate pathogens (Chaplin, 2010). Results from this study showed that PBMC from cattle identified as high immune responders had a significantly greater ability to proliferate compared to PBMC from low responders after HS1 and HS2 and had a trend towards greater cell proliferation after HS2 compared to average responders. Similarly, PBMC from average responders also had a trend towards greater proliferation compared to low responders after HS1 and HS2. Additionally, PBMC from high responders had greater cell proliferation compared to cells from low responders at TN. These results suggest that high responders are able to respond more effectively to invading pathogens both at TN conditions as well as after being exposed to a mild or moderate heat stress compared to low responders. Further, PBMC from average responders may generate a greater response to foreign pathogens compared low responders after single and multiple heat stress events. Peripheral Blood mononuclear cells from high responders trended toward increased cell proliferation after HS1 and HS2 compared to TN, whereas cells from average and low responders displayed similar ability to proliferate between TN, HS1 and HS2. As previously mentioned, studies have shown a decline in cell proliferation in relation to heat stress (Bhanuprakash et al., 2016), but others have found that cell proliferation was not affected by heat stress (Tao et al., 2011). Correspondingly, production of HSP has been linked to the ability of cells to proliferate under heat stress with the production of HSP being required for cell proliferation to occur (Pechan, 1991). This association between HSP production and cell proliferation was observed here, where both HSP70 concentration and cell proliferation increased in PBMC from high responders while there was no difference in HSP70 concentration and cell proliferation in PBMC from average and low responders.

The results observed here are in accordance with the results from a previous study in regards to cell proliferation of PBMC from immune phenotyped dairy cattle when exposed to *in-vitro* heat challenge (Cartwright et al., 2021). Both studies showed that high immune responding dairy cattle are likely to be more tolerant to mild or moderate heat stress compared to average and low responders. It has been shown that the negative effects of heat stress in dairy cattle, like reduced milk production, are associated with the diversion of energy to immune activation and inflammation (Gao et al., 2019; Kvidera et al., 2017). It is possible that some of the differences observed between immune phenotypes here is explained by differences in cellular energy metabolism, however, to date this has not been studied. It would be meaningful to investigate cell metabolism under TN and HS conditions to determine the role of cellular metabolism in differences observed here between immune phenotypes.

**CONCLUSIONS**

This *in-vivo* study corroborates the results observed in previous *in-vitro* heat stress studies of Holstein dairy cattle of diverse immune response phenotypes (Cartwright 2021). It also demonstrates the validity of the *in-vitro* heat challenge model using PBMC from dairy cattle as an *in-vivo* indicator of mild to moderate heat stress. Specifically, the results from this study illustrate that high immune responders appear to be more tolerant to a mild to moderate heat challenge as they demonstrate lower respiration rates at a higher THI and have superior cellular responses during mild to moderate heat stress as indicated by the production of greater concentrations of HSP70 and greater cell proliferation capability, compared to dairy cattle ranked as average or low responders. Overall, these results suggest that not only are high responders more resilient against disease, but they seem better able to tolerate mild to moderate heat stress, thus potentially reducing its impact on production and reproduction. Tolerance to heat stress in dairy cattle will be of increasing importance as climate change continues and atmospheric temperatures rise causing the frequency of heat events to increase and pathogen dynamics to change. This emphasizes the need for dairy cattle that are both more disease and heat tolerant. The results observed in this study provide evidence that high immune responding Holstein dairy cattle, already known for their superior immunity, are also likely more resilient to mild to moderate heat stress. Therefore, incorporating high responders in a dairy herd could aid in the prevention of economic losses from both disease and heat stress thereby improving animal welfare, all of which are essential as the climate continues to change.

## ACKNOWLEDGMENTS

The authors would like to thank the staff at the Elora Dairy Research Station, especially Laura Wright for their care of the dairy cattle used in this study and for ensuring experimental protocols were followed. The authors would also like to thank Marnie McKechnie, Siobhan Mellors and Royan Pariappaden for their technical support.

## FUNDING

The authors would like to acknowledge the support of the Canadian First Research Excellence Fund, Ontario Ministry of Agriculture and Rural Affairs and Dairy Farmer of Ontario.

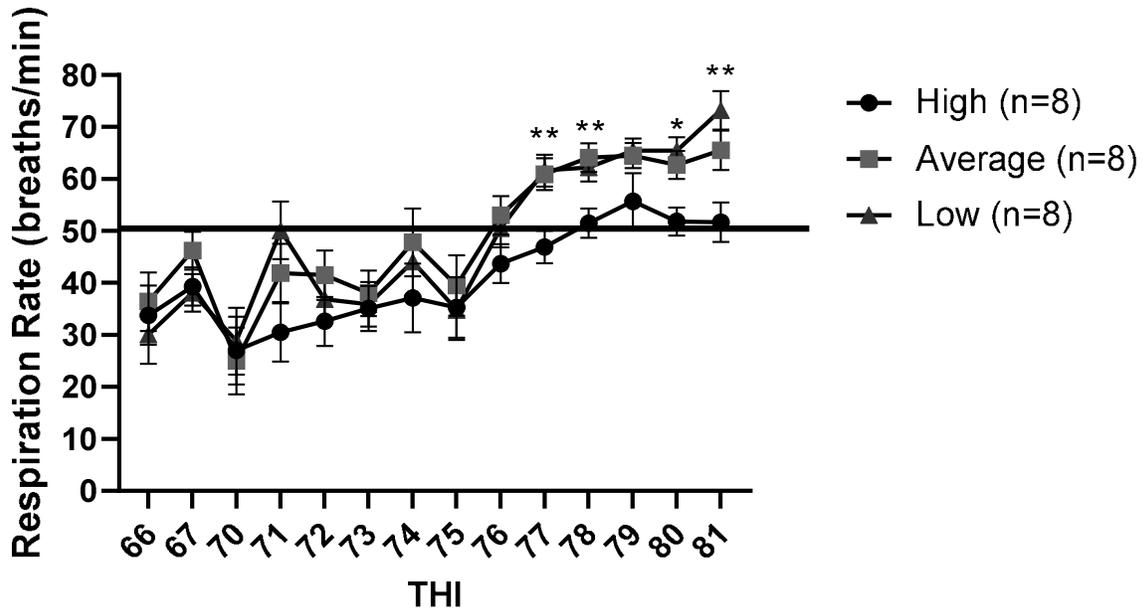

**Figure 1**: **Respiration rate in breaths per minute by temperature humidity index for high, average, and low immune responding dairy cattle**. The black line at 50 breaths per min represents the maximum normal respiration rate for dairy cattle (Becker et al., 2020). Each data point represents the LSM for each phenotype at that particular THI. Error bars represent SEM for LSM. Significance between high vs the other phenotypes (average and low) is represented by ** in the graph. Trends between high vs the other phenotypes (average and low) is represented by * in the graph. All other differences between phenotypes were not significant.

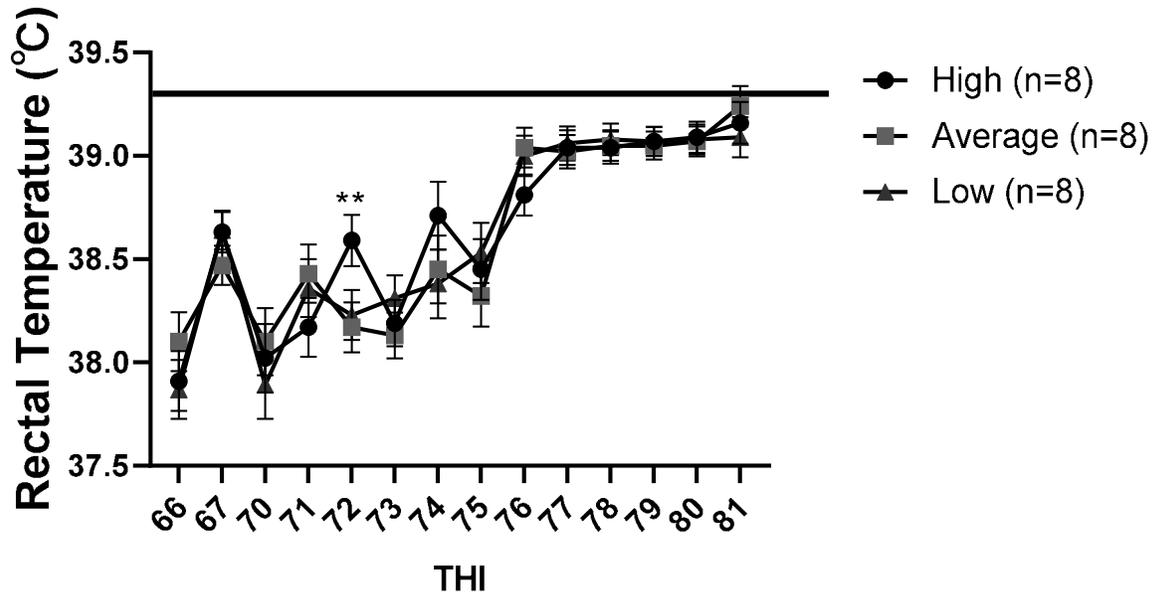

**Figure 2**: **Rectal temperature in °C by temperature humidity index for high, average, and low immune phenotyped dairy cattle**. The black like at 39.3°C represents the maximum normal rectal temperature in dairy cattle (Fielder, 2015). Each data point represents the LSM for each phenotype at that particular THI. Error bars represent SEM for LSM. Significance between high vs the other phenotypes (average and low) is represented by ** in the graph. All other differences between phenotypes were not significant.

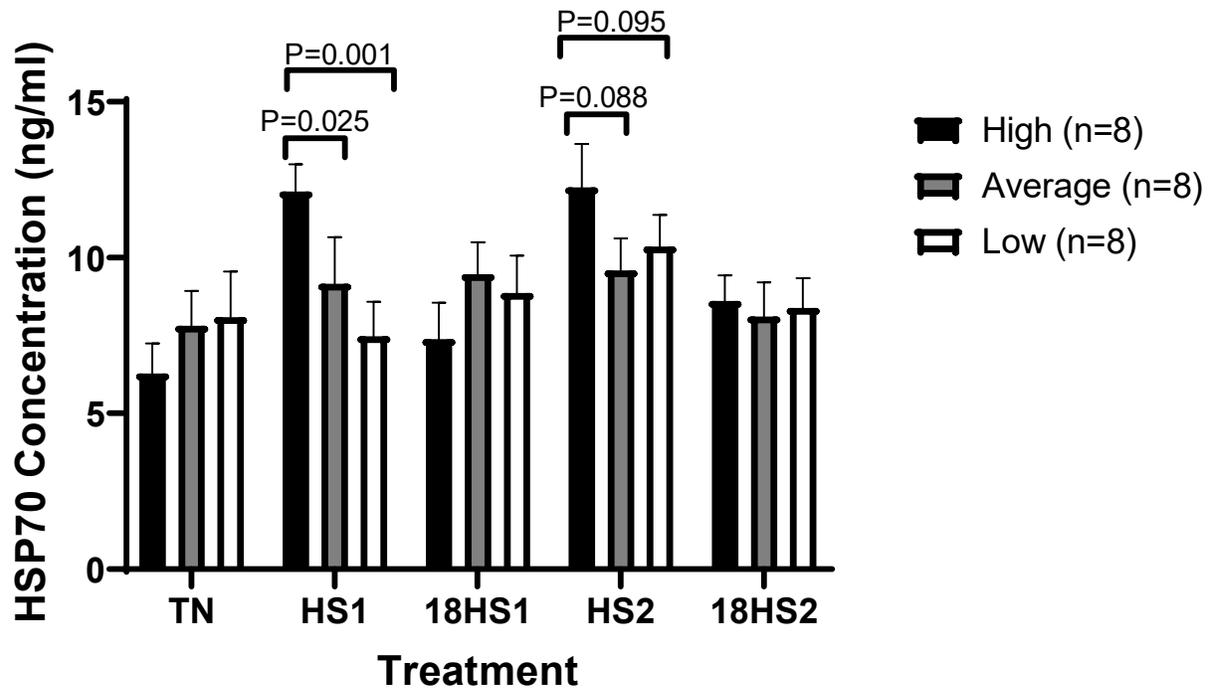

**Figure 3**: **HSP70 concentration in blood mononuclear cells after *in-vivo* heat challenge by immune response phenotype**. Each bar represents the LSM for each phenotype at that particular treatment. Error bars represent SEM for LSM. P-values for significance and trends between phenotypes are presented in the graph (HS1- High vs Average P = 0.025, High vs Low P = 0.001, Average vs Low P = NS, HS2- High vs Average P = 0.088, High vs Low P = 0.095, Average vs Low P = NS). If a p-value is not presented it means no significant difference was observed between phenotypes for that particular treatment.

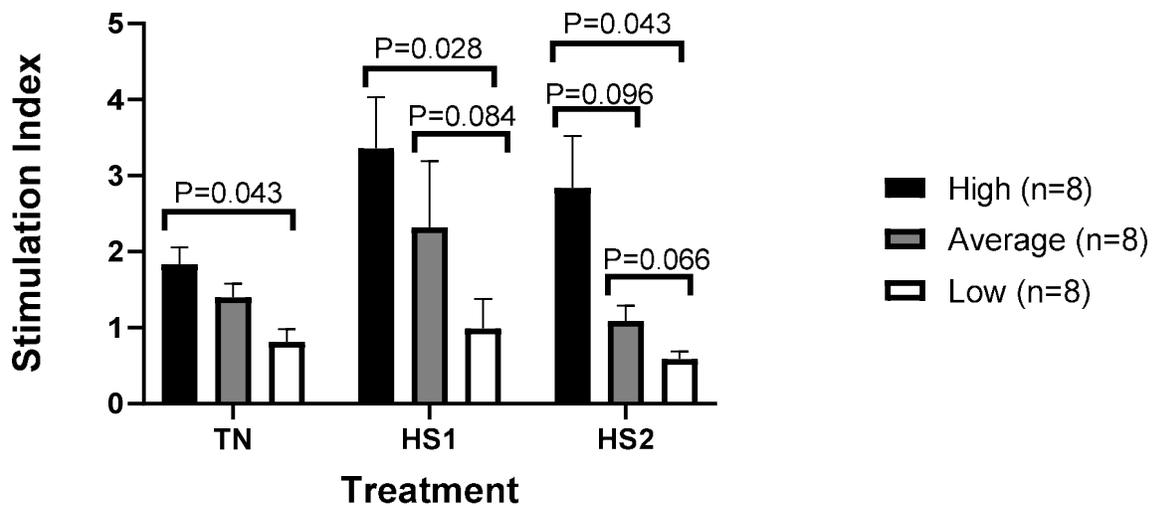

**Figure 4: Cell proliferation in blood mononuclear cells after *in-vivo* heat challenge by immune response phenotype**. Each bar represents the LSM for each phenotype at that particular treatment (TN = cells collected before the 1st heat challenge, HS1 = cells collected after cows exposed to one 4 hr heat challenge and HS2 = cells collected after cows exposed to two 4 hr heat challenges on subsequent days). Error bars represent SEM for LSM. P-values for significance and trends between phenotypes are presented in the graph (TN- High vs Low P = 0.043, HS1- High vs Low P = 0.028, Average vs Low P = 0.084, High vs Average P = NS, HS2 – High vs Average P = 0.096, High vs Low P = 0.043, Average vs Low P = 0.066). If a p-value is not presented it means no significant difference was observed between phenotypes for that particular treatment.